\documentstyle[12pt]{article}
\begin{document}
\title{Planck Scale Phenomena}
\author{B.G. Sidharth\\
Centre for Applicable Mathematics \& Computer Sciences\\
Adarsh Nagar, Hyderabad - 500 063, India}
\date{}
\maketitle
\begin{abstract}
Though the Planck scale is encountered in Quantum SuperString
Theory and Quantum Gravity, it is the Compton scale of elementary
particles which is encountered in the physical world. An
explanation for this is given in terms of Brownian processes and
the duality relation.
\end{abstract}
\section{Introduction}
It is well known that in Quantum Gravity as well as in Quantum
SuperString Theory, we encounter phenomena at the Planck scale.
Yet what we encounter in the real world is, not the Planck scale,
but the elementary particle Compton scale. The explanation for
this is that the very high energy Planck scale is moderated by the
Uncertainty Principle. The question which arises is, exactly how
does this happen? We will now present an argument to show how the
Planck scale leads to the real world Compton scale, via
fluctuations and a modification of the Uncertainty Principle.
\section{The Planck Scale}
It is well known that the Planck Scale is defined by
$$l_P = \left(\frac{\hbar G}{c^3}\right)^{\frac{1}{2}} \sim
10^{-33}cm$$
\begin{equation}
t_P = \left(\frac{\hbar G}{c^5}\right)^{\frac{1}{2}} \sim
10^{-42}sec\label{e1}
\end{equation}
(\ref{e1}) defines the absolute minimum physical scale
\cite{r1,r2}. Associated with (\ref{e1}) is the Planck mass
\begin{equation}
m_P \sim 10^{-5}gm\label{e2}
\end{equation}
There are certain interesting properties associated with
(\ref{e1}) and (\ref{e2}). $l_P$ is the Schwarzschild radius of a
black hole of mass $m_P$ while $t_P$ is the evaporation time for
such a black hole via the Beckenstein radiation \cite{r3}.
Interestingly $t_P$ is also the Compton time for the Planck mass,
a circumstance that is symptomatic of the fact that at this scale,
electromagnetism and gravitation become of the same order
\cite{r4}. Indeed all this fits in very well with Rosen's analysis
that such a Planck scale particle would be a mini universe
\cite{r5,r6}. We will now invoke a time varying gravitational
constant
\begin{equation}
G \approx \frac{lc^2}{m\sqrt{N}} = (\sqrt{N}t)^{-1} \alpha
T^{-1}\label{e3}
\end{equation}
which resembles the Dirac cosmology and features in another scheme
in which (\ref{e3}) arises due to the fluctuation in the particle
number \cite{r7,r8,r9,r10,r4}. In (\ref{e3}) $m$ and $l$ are the
mass and Compton wavelength of a typical elementary particle like
the pion while $N \sim 10^{80}$ is the number of elementary
particles in the universe, $\sqrt{N}$ the fluctuation in particle
number and
$T$ is the age of the universe.\\
In this scheme wherein (\ref{e3}) follows from the theory, we use
the fact that given $N$ particles, the fluctuation in the particle
number is of the order $\sqrt{N}$, as noted by Hayakawa, while a
typical time interval for the fluctuations is $\sim \hbar /mc^2$,
the Compton time. We will come back to this point later. So we
have
$$\frac{dN}{dt} = \frac{\sqrt{N}}{\tau}$$
whence on integration we get,
$$T = \frac{\hbar}{mc^2} \sqrt{N}$$
and we can also deduce its spatial counterpart, $R = \sqrt{N} l$,
which is the well known empirical Eddington formula. It is also
possible to then deduce the Hubble Law and a hitherto mysterious
empirical relationship, noted by Weinberg, between the mass of the
pion and the Hubble
constant (Cf.refs. for details).\\
Equation (\ref{e3}) which is an order of magnitude relation is
consistent with observation \cite{r11} while it may be remarked
that the Dirac cosmology itself has inconsistencies.\\
Substitution of (\ref{e3}) in (\ref{e1}) yields
$$l = N^{\frac{1}{4}} l_P,$$
\begin{equation}
t = N^{\frac{1}{4}} t_P\label{e4}
\end{equation}
where $t$ as noted is the typical Compton time of an elementary
particle. We can easily verify that (\ref{e4}) is consistent. It
must be stressed that (\ref{e4}) is not a fortuitous empirical
coincidence, but rather is a result of using (\ref{e3}), which
again as noted, can be deduced from
theory.\\
(\ref{e4}) can be rewritten as
$$l = \sqrt{n}l_P$$
\begin{equation}
t = \sqrt{n}t_P\label{e5}
\end{equation}
wherein we have used (\ref{e3}) and $n = \sqrt{N}$.\\
We will now compare (\ref{e5}) with the well known relations,
deduced earlier,
\begin{equation}
R = \sqrt{N} l \quad T = \sqrt{N} t\label{e6}
\end{equation}
The first relation of (\ref{e6}) is the well known Eddington
formula referred to while the second relation of (\ref{e6}) is
given also on the right side of (\ref{e3}). We now observe that
(\ref{e6}) can be seen to be the result of a Brownian Walk
process, $l,t$ being typical intervals between "steps"
(Cf.\cite{r4,r12,r13}). We demonstrate this below after equation
(\ref{e8}). On the other hand, the typical intervals $l,t$ can be
seen to result from a Nelsonian process themselves. Let us
consider Nelson's relation,
\begin{equation}
(\Delta x)^2 \equiv l^2 = \frac{\hbar}{m} t \equiv \frac{\hbar}{m}
\Delta t\label{e7}
\end{equation}
(Cf.\cite{r14,r15,r16,r17,r12}).\\
Indeed as $l$ is the Compton wavelength, (\ref{e7}) can be
rewritten as the Quantum Mechanical Uncertainty Principle
$$l \cdot p \sim \hbar$$
at the Compton scale (Cf. also \cite{r18}) (or even at the de
Broglie
scale).\\
What (\ref{e7}) shows is that a Brownian-Nelsonian process defines
the Compton scale while (\ref{e6}) shows that a Random Walk
process with the Compton scale as the interval defines the length
and time scales of the universe itself (Cf.\cite{r13}). Returning
now to (\ref{e5}), on using (\ref{e2}), we observe that in
complete analogy with (\ref{e7}) we have the relation
\begin{equation}
(\Delta x)^2 \equiv l^2_P = \frac{\hbar}{m_P} t_P \equiv
\frac{\hbar}{m_P} \Delta t\label{e8}
\end{equation}
We can now argue that the Nelsonian-Brownian process (\ref{e8})
defines the Planck length while a Brownian Random Walk process
with the Planck scale as the interval leads to (\ref{e5}), that is
the
Compton scale.\\
To see all this in greater detail, it may be observed that
equation (\ref{e8}) (without subscripts)
\begin{equation}
(\Delta x)^2 = \frac{\hbar}{m} \Delta t\label{ea}
\end{equation}
is the same as the Nelsonian equation, indicative of a double
Weiner process. Indeed as noted by several scholars, this defines
the fractal Quantum path of
dimension 2 (rather than dimension 1) (Cf.e.g. ref.\cite{r15}).\\
Firstly it must be pointed out that equation (\ref{ea}) defines a
minimum space time unit - the Compton scale $(l,t)$. This follows
from (\ref{ea}) if we substitute into it $\langle \frac{\Delta x}
{\Delta t}\rangle_{max} = c$. If the mass of the particle is the
Planck mass, then this Compton scale becomes the Planck scale.\\
Let us now consider the distance traversed by a particle with the
speed of light through the time interval $T$. The distance $R$
covered would be
\begin{equation}
\int dx = R = c \int dt = cT\label{eI}
\end{equation}
by conventional reasoning. In view of the Nelsonian equation
(\ref{ea}), however we would have to consider firstly, the minimum
time interval $t$ (Compton or Planck time), so that we have
\begin{equation}
\int dt \to nt\label{eII}
\end{equation}
Secondly, because the square of the space interval $\Delta x$
(rather than the interval $\Delta x$ itself as in conventional
theory) appears in (\ref{ea}), the left side of (\ref{eI})
becomes, on using (\ref{eII})
\begin{equation}
\int dx^2 \int (\sqrt{n}dx) (\sqrt{n}dy)\label{eIII}
\end{equation}
Whence for the linear dimension $R$ we would have
\begin{equation}
\sqrt{n}R = nct \quad \mbox{or} \quad R = \sqrt{n} l\label{eIV}
\end{equation}
Equation (\ref{eIII}) brings out precisely the fractal dimension
$D = 2$ of the Brownian path while (\ref{eIV}) is identical to
(\ref{e4}) or (\ref{e6}) (depending on whether we are dealing with
minimum intervals of the Planck scale or Compton scale of
elementary particles). Apart from showing the Brownian character
linking equations (\ref{e4}) and (\ref{ea}), incidentally, this
also provides the justification for what has so far been
considered to be a mysterious large number coincidences viz. the
Eddington
formula (\ref{e6}).\\
There is another way of looking at this. It is well known that in
Quantum SuperString Theory, at the Planck scale we have a non
commutative geometry \cite{r19,r20}
\begin{equation}
[x,y] \approx 0 (l^2_P), [x,p_x] = \hbar [1 +
0(l^2_P)]etc.\label{e9}
\end{equation}
Indeed (\ref{e9}) follows without recourse to Quantum
SuperStrings, merely by the fact that $l_P,t_P$ are the absolute
minimum space
time intervals as shown a long time ago by Snyder \cite{r21}.\\
The non commutative geometry (\ref{e9}), as is known is
symptomatic of a modified uncertainty principle at this scale
\cite{r22}-\cite{r28}
\begin{equation}
\Delta x \approx \frac{\hbar}{\Delta p} + l^2_P \frac{\Delta
p}{\hbar}\label{e10}
\end{equation}
The relation (\ref{e10}) would be true even in Quantum Gravity.
The extra or second term on the right side of (\ref{e10})
expresses the well known duality effect - as we attempt to go down
to the Planck scale, infact we are lead to the larger scale
represented by it. The question is, what is this larger scale? If
we now use the fact that $\sqrt{n}$ is the fluctuation in the
number of Planck particles (exactly as $sqrt{N}$ was the
fluctuation in the particle number as in (\ref{e3}) so that
$\sqrt{n}mpc = \Delta p$ is the fluctuation or uncertainty in the
momentum for the second term on the right side of (\ref{e10}), we
obtain for the uncertainty in length,
\begin{equation}
\Delta x = l^2_P \frac{\sqrt{n}m_Pc}{\hbar} =
l_P\sqrt{n},\label{e11}
\end{equation}
We can easily see that (\ref{e11}) is the same as the first
relation of (\ref{e5}). The second relation of (\ref{e5}) follows
from an
application of the time analogue of (\ref{e10}).\\
Thus the impossibility of going down to the Planck scale because
of (\ref{e9}) or (\ref{e10}), manifests itself in the fact that as
we attempt to go down to the Planck scale, we infact end up at the
Compton scale.\\
Interestingly while at the Planck length, we have a left time of
the order of the Planck time, as noted above it is possible to
argue on the other hand that with the pion mass and length of a
typical elementary particle like the pion, at the Compton scale,
we have a life time which is the age of the universe itself as
shown by
Sivaram \cite{r3,r29}.\\
Interestingly also Ng and Van Dam deduce the relations like
\cite{r30}
\begin{equation}
\delta L \leq (Ll^2_P)^{1/3}, \delta T \leq
(Tt^2_P)^{1/3}\label{e12}
\end{equation}
where the left side represents the uncertainty in the measurement
of length and time for an interval $L,T$. We would like to point
out that if in (\ref{e12}) we use for $L,T$, the size and age of
the universe, then $\Delta L$ and
$\Delta T$ reduce to the Compton scale $l,t$.\\
In conclusion, Brownian-Nelsonian processes and the modification
of the Uncertainity Principle at the Planck scale lead to the
physical Compton scale.

\end{document}